\documentclass[twocolumn,10pt,aps,prd,nofootinbib,floatfix]{revtex4-1}

\usepackage{amssymb,amsmath,amsthm}
\usepackage[pdftex]{graphicx} 
\usepackage{braket}
\usepackage[T1]{fontenc}
\usepackage[latin1]{inputenc}
\usepackage[english]{babel}

\usepackage{float}
\usepackage{braket}
\usepackage{slashed}
\usepackage{commath}
\usepackage{diagbox}
\usepackage{color}
\usepackage{bm}
\usepackage[breaklinks=true]{hyperref} 
\hypersetup{colorlinks=true,citecolor=blue,linkcolor=blue,urlcolor=blue}
\usepackage{booktabs}
\usepackage{mathrsfs}
\usepackage{enumitem}

\addto\captionsenglish{}

\addto\captionsenglish{}

\renewcommand{\epsilon}{\varepsilon}
\newcommand{\mb}[0]{\mathbf}

\renewcommand{\rm}[1]{\mathrm{#1}}

\newcommand{\w}{w^{\rm{add}}}

\newcommand{\psibar}{\overline{\psi}}
\newcommand{\ubar}{\bar{u}}

\newcommand{\HankelH}{H^{(2)}}

\begin{document}
\title[]{Fermionic decay of a massive scalar in the early Universe}

\author{Juho Lankinen}
\email{jumila@utu.fi}
\affiliation{Turku Center for Quantum Physics, Department of Physics and Astronomy, University of Turku, Turku 20014, Finland}

\author{Joonas Malmi}
\email{joonas.e.malmi@utu.fi}
\affiliation{Turku Center for Quantum Physics, Department of Physics and Astronomy, University of Turku, Turku 20014, Finland}

\author{Iiro Vilja}
\email{vilja@utu.fi}
\affiliation{Turku Center for Quantum Physics, Department of Physics and Astronomy, University of Turku, Turku 20014, Finland}

\begin{abstract}
We derive a curved space generalization of a scalar to fermion decay rate with a Yukawa coupling in expanding Friedmann-Robertson-Walker universes. This is done using the full theory of quantum fields in curved spacetime and the added-up transition probability method. It is found that in  an expanding universe the usual Minkowskian decay rates are considerably modified for early times. For conformally coupled scalars the decay rate is modified by a positive additive term proportional to the inverse of mass and related to the expansion rate of the Universe. We compare and contrast our results with previous studies on scalar to scalar decay and find that in general the decay channel into fermions is the dominant channel of decay in the very early Universe.

\end{abstract}

\maketitle

\section{Introduction}

Particle decay processes in the early Universe have deep and profound implications in cosmology, from baryogenesis \citep{Kolb_Riotto_Tkachev:1998,Babu_Mohapatra:2012} to Big Bang nucleosynthesis \citep{Cyburt:2016} and reheating scenarios after inflation \citep{Allahverdi:2010}. When considering these early Universe processes in depth, the influence of spacetime curvature cannot be neglected anymore. In this regime the Minkowskian quantum field theory is ultimately only an approximation and of limited applicability. Instead, when studying particle processes in the early Universe, or in any kind of situation where the effects of gravitation cannot be neglected, quantum field theory in curved spacetime must be used. As a result of spacetime curvature and the related violation of conservation laws, the particle decay rates, cross sections and lifetimes are modified compared to usual flat space results \citep{Audretsch_Spangehl:1985,Lankinen_Vilja:2018b}. New particle processes, forbidden in Minkowski space, are to be considered leading to new Feynman diagrams even at first-order \citep{Audretsch_Spangehl:1985,Lotze:1989a,Lotze:1989b}. The lack of energy conservation, e.g., in Friedmann-Robertson-Walker (FRW) universes, leads to processes where a particle can even decay into quanta of its own field \citep{Boyanovsky_Holman_Kumar:1997,Boyanovsky_deVega:2004}. All these suggest that considerable care must be exercised when studying mutually interacting fields in curved spacetime.

The groundwork for interacting quantum fields in expanding universes was laid in the 1980s expanding the familiar $S$ matrix formulation of in-out states to simple cosmological models. Studies focused on renormalization of self-interacting fields \citep{Birrell_Ford:1979} but also on massive scalar particles decaying into two conformally coupled scalar particles \citep{Audretsch_Spangehl:1985,Audretsch_Ruger_Spangehl:1987,Audretsch_Spangehl:1986,Audretsch_Spangehl:1987}. The method known as added-up probability, introduced in \citep{Audretsch_Spangehl:1985}, was used to study photon decay in radiation dominated universe \citep{Tsaregorodtsev:2004} and more recently in investigating scalar particle decay in expanding FRW  universes \citep{Lankinen_Vilja:2017b,Lankinen_Vilja:2018a,Lankinen_Vilja:2018b}. These investigations into scalar decay revealed that the Minkowskian decay rates are significantly modified at early times, but in the long-time limit approach the Minkowskian results asymptotically. Lately, a nonperturbative method, the so called Wigner-Weisskopf method \citep{Boyanovsky_Holman:2011} familiar from quantum optics, was adopted to study particle decay in the context of inflationary cosmology \citep{Lello_Boyanovsky_Holman:2013} as well as post inflationary cosmology \citep{Herring:2018} and to study decay in renormalizable field theories \citep{Boyanovsky:2019}.

Although these studies have to a great degree increased our knowledge of particle decay in curved spacetime, they have mainly been concerned with a scalar channel decay, see e.g., \cite{Audretsch_Spangehl:1985,Audretsch_Ruger_Spangehl:1987,Lankinen_Vilja:2017b,Lankinen_Vilja:2018a,Lankinen_Vilja:2018b}. The negligence of the fermionic channel may be attributed to the more complex nature of the problem involving spinors which may lead to some mathematical difficulties encountered in the technical calculations. Clearly, the study of the fermionic decay channel would further increase and complete our knowledge of particle decay in curved space. This serves as the motivation for this paper where we will consider a scalar field interacting with a Dirac spinor through a Yukawa interaction in curved space. By applying the added-up method, we  calculate the decay rate for a fermionic decay of a massive scalar in an expanding spatially flat FRW universe. The scalar field is taken with an arbitrary coupling $\xi$ to gravity hence avoiding the restrictions commonly encountered in literature where $\xi$ is usually restricted to be that of minimally or conformally coupled. The fermions are assumed to be massless in order to to avoid interpretational issues of the decay rate when using the added-up method.

This article is structured in the following way. In Sec. \ref{Sec:2} we introduce the necessary theoretical background and give a brief review of the method of added-up probability. In Sec. \ref{Sec:3} we obtain the positive mode solutions for both the spinor field and the scalar field. These modes are used in Sec. \ref{Sec:4} to derive a transition amplitude in curved space for the scalar to decay into a fermion-antifermion pair. In this section we will also show that the transition amplitude has the correct Minkowskian limit. After that we will compare the decay into fermions with the recently obtained decay into massless scalars in Sec. \ref{Sec:5} where it is shown that the fermionic decay channel is dominant to the scalar channel in the early Universe. Finally, we will discuss the implications and reasons behind our results in Sec. \ref{Sec:6} and end with conclusions in Sec. \ref{Sec:7}.

\section{Preliminaries}\label{Sec:2}
As was stated in the introduction, there are at least two distinct ways to calculate the decay probability in curved spacetime. The difference between these two methods lies in the quantity which is eventually being calculated. The added-up method is fundamentally a global expression for the decay rate, while the Wigner-Weisskopf method is concerned more with the differential decay rate. In this article we will use the added-up method introduced in \citep{Audretsch_Spangehl:1985} and the purpose of this section is to summarize this method. We begin by introducing the necessary theoretical background and then give a brief review of the method of added-up probability. Natural units $\hbar=c=1$ are used and the metric is chosen with a positive time component.

\subsection{Theoretical background}
We will consider a model described by a four-dimensional spatially flat Friedmann-Robertson-Walker spacetime with the metric
 	\begin{align}
	ds^2=dt^2-a(t)^2 d\mathbf{x}^2
	\end{align}
given in standard coordinate time $t$ with a dimensionless scale factor $a(t)$. On this classical curved background we consider propagating quantum fields mutually interacting with each other. In particular we consider a Yukawa type interaction where a massive scalar particle decays into two massless fermions.  
The Lagrangian density $\mathcal{L}$ for this theory is given by
	\begin{align}
	\mathcal{L}=\mathcal{L}_\phi + \mathcal{L}_\psi + \mathcal{L}_I,
	\end{align}
which consists of a Lagrangian density for the complex scalar field $\mathcal{L}_\phi$, a fermion Lagrangian $\mathcal{L}_\psi$ and an interaction term $\mathcal{L}_I$.
	
For a complex scalar field in curved space the Lagrangian density is straightforwardly given as
	\begin{align}
	\mathcal{L}_\phi=\sqrt{-g}(\partial_\mu\phi^* \partial^\mu\phi-m^2\phi^*\phi-\xi R\phi^*\phi),
	\end{align}		
where $g$ is the determinant of the metric, $R$ the Ricci scalar and the coupling of the scalar field to gravity is controlled by the dimensionless parameter $\xi$. In four dimensions, the value $\xi=1/6$ is known as conformal coupling, while the minimal coupling is given by $\xi=0$. Denoting the covariant d'Alembert operator by $\square$, the Klein-Gordon equation in curved spacetime for the scalar field $\phi$ is given by
	\begin{align}\label{eq:KGCurved}
	(\square+m^2+\xi R)\phi=0.
	\end{align}

The fermion part of the Lagrangian requires a little more thought in curved space. Because the spinor does not transform like a tensor under Lorentz transformation, one cannot just replace the derivatives with their covariant counterparts. To incorporate spinors into general relativity, one can introduce a set of four covariant vector fields $e_a{}^\mu$ known as a tetrad. We adopt a convention where the latin indices refer to local inertial coordinates while greek indices refer to general coordinates. With this formalism, the Lagrangian for a massless spinor field in curved spacetime is  given by
	\begin{align}
	\mathcal{L}_\psi=\frac{i}{2}\sqrt{-g}(\psibar \gamma^\mu\nabla_\mu \psi-(\nabla_\mu\psibar) \gamma^\mu \psi),
	\end{align}
where $\psibar$ denotes the Dirac conjugate spinor $\psibar=\gamma^0\psi^\dagger$. The curved space gamma matrices are defined via the tetrad as $\gamma^\mu=e_a{}^\mu \gamma^a$, where $\gamma^a$ denotes the usual flat space gamma matrix. The curved space gamma matrices satisfy the usual anticommutation relations
	\begin{align}
	\{\gamma^\mu, \gamma^\nu\}=2g^{\mu\nu}
	\end{align}	  
and the covariant derivative is defined with the help of a spin-connection $\Gamma_\mu$ as
	\begin{align}
	\nabla_\mu:=\partial_\mu+\Gamma_\mu,
	\end{align}
where
	\begin{align}
	\Gamma_\mu=\frac{1}{8}[\gamma_a,\gamma_b]e_a{}^\nu \partial_\mu e_{b\nu}.
	\end{align}
By varying the action one obtains the Dirac equation for a massless $\psi$-particle in curved space
	\begin{align}\label{eq:DiracCurved}
	i\gamma^\mu \nabla_\mu\psi=0.
	\end{align}
	
The choice of a massless spinor field is motivated by its necessity in the added-up method, but this choice can also be thought from a perspective of symmetry. If one considers a complex scalar field with  non-zero charge, a chiral spinor field $\psi_L$ with opposite charge and  a zero-charge field $\psi_R$, in a $U(1)$-symmetric theory a mass term in the fermion Lagrangian density violates this symmetry while the scalar Lagrangian density is left unchanged. It should also be noted that in this case no mass appears in the renormalization procedure either. Therefore, if one wants to consider a (globally) $U(1)$-symmetric theory with chiral fields, then the fermionic Lagrangian density should be taken as massless. Moreover, if the field $\phi$ is considered as an $SU(2)$ doublet, by defining the (global) $U(1)$-charges of the theory in a suitable way, the fermionic fields may be thought of as fermions of the Standard Model.

Because a complex scalar field can always be decomposed into two real scalar fields as $\phi=(\phi_1+i\phi_2)/\sqrt{2}$, from now on we will consider only real scalar fields. This is merely to make the technical calculations and theory more manageable. In what follows, the only effect of this onto the calculations is that the total transition amplitude is to be multiplied by a factor of two in the complex case. For a Yukawa type interaction the interaction part of the Lagrangian when considering real scalar fields is of the form
	\begin{align}
	\mathcal{L}_I=-\sqrt{-g}h\psibar \phi \psi,
	\end{align}
where $h\neq 0$ is a dimensionless coupling constant chosen to be real.  To describe the interaction, we will apply the $S$ matrix scheme where the $S$ matrix is given as
	\begin{align}
	S=\lim_{\beta\to 0^+}\hat{T}\exp\Big(-i\int\sqrt{-g}e^{-\beta t}h \psibar \phi \psi d^4x\Big),
	\end{align}
where $\hat{T}$ denotes the time-ordering operator. The exponential factor $e^{-\beta t}$ acts as a switch-off for the interaction for large times with $\beta$ being a positive constant and called the switch-off parameter. The perturbative expansion of the $S$ matrix for this interaction gives
	\begin{align}
	S=1-ih A+\mathcal{O}(h^2),
	\end{align}
where
	\begin{align}\label{A_integraali}
	A:=\lim_{\beta\to 0^+}\int \hat{T}\psibar \phi \psi e^{-\beta t}\sqrt{-g}\, d^4x.
	\end{align}
Furthermore, we consider only tree level processes for which the transition amplitude is defined as 
	\begin{align}
\mathscr{A}:= \braket{\rm{out}|A|\rm{in}}.
	\end{align}

\subsection{The added-up probability}
Recall a result concerning free fields in curved spacetime: conformally coupled massless particles are not created as a result of spacetime expansion \citep{Parker:1969,Parker:1971}. This result, which applies both for spin-$0$ as well as spin-$1/2$ particles, is at the heart of the added-up probability method introduced in \citep{Audretsch_Spangehl:1985}. Because of this result, a detection of a massless conformally coupled particle in the asymptotic region of spacetime indicates that this particle must have been solely created or influenced by a decay process. 

\begin{figure}[H]
	\centering
	\includegraphics[scale=1]{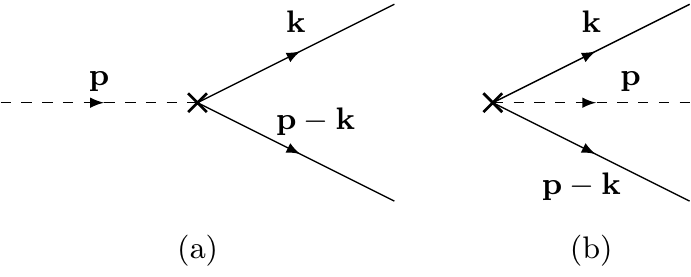}
	\caption{Production of massless fermions (a) with and (b) without  proper decaying massive state. The 		dashed line corresponds to the massive scalar and the solid lines to massless fermions. The vertex cross  indicates gravitational influence.}\label{fig:1} 
	\end{figure}

In curved spacetime there is still one other complication arising from the lack of energy conservation in curved space; the creation of all three particles with a priori unrestricted momenta. This lack of kinematic thresholds was also considered in \citep{Boyanovsky_Holman:2011} using the Wigner-Weisskopf method.
Because of this, it is necessary in the added-up method to further restrict the out momenta to fulfill the three-momentum conservation law $\mathbf{p}=\mb k_1+\mb k_2$. Only after that one is left with what resembles closest a decay process in curved space. This process can be described at tree level with two Feynman diagrams (Fig.\ref{fig:1}). 
	
In the original treatment of Audretsch and Spangehl only decay into scalar particles was concerned. Nevertheless, the formalism used in the added-up method applies equally well for other types of decay product particles as long as they are massless and conformally coupled to facilitate a meaningful decay rate interpretation in curved space. Especially this also holds for the case where a particle decays into two conformally coupled massless fermions. For a more exhaustive treatment of the added-up method, we refer the reader to the original paper \citep{Audretsch_Spangehl:1985}.  

The added-up transition amplitude corresponding to the two Feynman diagrams in Fig. \ref{fig:1} reads as
	\begin{align}\nonumber\label{eq:w_add}
	w^{\rm{add}}(\mb p,\mb k,\mb{p-k}) =& h^2 \Big\{ \lvert \braket{\rm{out}, 1^{\psibar}_{\mb{k}} 1^\psi_{\mb{p-k}}\lvert A \rvert 1^\phi_{\mb{p}}, \rm{out}}\rvert^2\\
	&+\lvert\braket{\rm{out},1^\phi_{\mb{-p}} 1^{\psibar}_{\mb{k}}1^\psi_{\mb{p-k}}\lvert A \rvert 0, \rm{out}}\rvert^2 \Big\},
	\end{align}
where $\mb k_1=\mb k$ and $\mb k_2=\mb{p-k}$ and $\ket{\mathrm{out}}$ refers to the state in the out region, i.e., the asymptotic future. In the formalism one could use both the in or out regions for the one-particle and vacuum states but not the usual in and out states used in flat space. This is because such a matrix element suffers from physical interpretational issues in curved spacetime \citep{Audretsch_Spangehl:1985}. With this in mind we choose the asymptotic out region for the simple reason that the massive field modes can be recognized in the out region.  The total transition probability $w^{\mathrm{tot}}$ is obtained by summing over all the $k$ modes,
	\begin{align}\label{eq:w_tot}
	w^{\mathrm{tot}}=\sum_{\mb k} \w(\mb p,\mb k,\mb{p-k}).
	\end{align}	
	
With this brief review of the added-up method we turn our attention to finding the positive field modes needed in the calculations.

\section{The positive mode solutions}\label{Sec:3}
One of the most important issues in quantum field theory in curved spacetime is finding analytic solutions to the field equations. This is a task of great complexity because in general one is lead to solving highly non-trivial differential equations. Even when the solution is obtained, it is not all that certain that the positive mode solutions can be obtained. In any case, one must fix the scale factor in order to even begin finding the field modes.

The literature is filled with numerous studies for finding scalar field modes for different types of universes, see e.g., \citep{Birrell_Davies,Parker_Toms} and references therein. Many models studied are non-physical with the purpose of merely illustrating the effects of particle creation from the vacuum. Even if a positive field mode can be found for a more general realistic model, it usually suffers from quite strict restrictions \citep{Lankinen_Vilja:2018b}.
The fermionic case presents no simplifications into finding the positive modes. Some exact solutions have been reported \citep{Barut_Duru:1987,Villalba_Percoco:1991,Chimento_Mollerach:1986} and also more recently by Moradi \citep{Moradi:2009,Moradi:2008}. The goal of this section is to construct the positive field modes for the spinor and scalar fields used in the calculation of the transition probability in the next section.

\subsection{Spinor modes}	
As noted, finding a solution to the curved space Dirac equation is a highly technical task and generally one needs to fix the scale factor explicitly in order to obtain an exact solution. This is the case even when considering a massless Dirac equation like Eq. \eqref{eq:DiracCurved}. In a spatially flat FRW universe, the most general scale factor we can choose in this paper is a typical power-law expansion
	\begin{align}
	a(t)=bt^n,
	\end{align}
where $b$ is a positive constant controlling the expansion rate of the Universe. We take $n \in [0,1)$ which covers most of the interesting cosmological cases like radiation and matter dominated universes. The de Sitter solution is excluded in this paper because a positive mode solution for the massive scalar field cannot be found for de Sitter space \citep{Lankinen_Vilja:2018b}.
 
Let us begin by expanding the curved space Dirac equation \eqref{eq:DiracCurved} in flat FRW spacetime. With the substitution of $\psi=a^{-3/2}\tilde{\psi}$ we can further simplify the ensuing equation to obtain
	\begin{align}\label{eq:DiracReduced}
	i\gamma^0 \tilde{\psi}'(t)-\frac{1}{a(t)}\gamma^i k_i \tilde{\psi}(t)=0.
	\end{align}	 
Here $k_i$ denotes the vector component of the momentum vector $\mb k$ and $\gamma^a$ are given in chiral representation as
	\begin{align}
	\gamma^0=\begin{pmatrix}
		 0 && \sigma_0\\
		 \sigma_0 && 0\\
	     \end{pmatrix},
	     \
	     \gamma^i=\begin{pmatrix}
		 0 && \sigma_i\\
		 -\sigma_i && 0\\
	     \end{pmatrix},
\end{align}	 
where $\sigma_0$ is the identity matrix and $\sigma_i$ are the Pauli matrices. The advantage of using chiral representation is that Eq. \eqref{eq:DiracReduced} reduces upon solving e.g., the spinor components $\psi_2,\psi_4$ into sets of two second-order differential equations
	\begin{align}
	\tilde{\psi}''_i+\frac{\dot{a}}{a}\tilde{\psi}'_i+\frac{|\mb k|^2}{a^2}\tilde{\psi}_i=0, \ i=1,3,
	\end{align}
for the components $\psi_1$ and $\psi_3$. This differential equation can readily be solved and upon substituting the answers to the remaining equations for $\psi_2$ and $\psi_4$ obtained from Eq. \eqref{eq:DiracReduced} yields the unnormalized solutions for the spinor components. From the solutions we obtain a set of mode solutions for the massless fermion in chiral representation,
\begin{align}
	u_{\mb k}^s(t,\mb x)&=\frac{1}{[2\pi a(t)]^{3/2}\sqrt{2k}}u(\mb k,s)e^{i\mb{k\cdot x}-\frac{ik}{b(1-n)}t^{1-n}}\\
	v_{\mb k}^s(t,\mb x)&=\frac{1}{[2\pi a(t)]^{3/2}\sqrt{2k}}v(\mb k,s)e^{-i\mb{k\cdot x}+\frac{ik}{b(1-n)}t^{1-n}},
\end{align}	
where the positive and negative energy spinors are to be normalized according to
	\begin{align}
	u(\mb k,s)^\dagger u(\mb k,s')=v(\mb k,s)^\dagger v(\mb k,s')=2|\mb k|\delta_{s s'}.
	\end{align}
Applying this normalization condition the normalized modes are given as
\begin{align}\nonumber
	u(\mb k,+)&= \begin{pmatrix}
		 \sqrt{k-k_3}\\
		 \frac{-k_+}{\sqrt{k-k_3}}\\
		 0\\
		 0	
	     \end{pmatrix}, \quad
	v(\mb k,+)= \begin{pmatrix}
		 \sqrt{k+k_3}\\
		 \frac{-k_+}{\sqrt{k-k_3}}\\
		 0\\
		 0	
	     \end{pmatrix},\\
	u(\mb k,-)&= \begin{pmatrix}
		 0\\
		 0\\
		 \sqrt{k+k_3}\\
		 \frac{k_+}{\sqrt{k+k_3}}	
	     \end{pmatrix},	  \quad
	v(\mb k,-)= \begin{pmatrix}
		 0\\
		 0\\
		 \sqrt{k-k_3}\\
		 \frac{k_+}{\sqrt{k+k_3}}	
	     \end{pmatrix}	 	        ,
\end{align}
where we have defined $k:=|\mb k| $ and $k_{\pm}:=k_1 \pm i k_2$. The fields can thus be expanded as
	\begin{align}
	\psi&= \sum_{\pm s}\int d^3\mb k\big[ b_{\mathbf{k}}^s u_{\mathbf{k}}^s(t,\mathbf{x})  +  d_{\mathbf{k}}^{s^\dagger} v_{\mathbf{k}}^s(t,\mathbf{x}) \big]\\
	\psibar&=\sum_{\pm s}\int d^3\mb k \big[ d_{\mathbf{k}}^s \bar{v}_{\mathbf{k}}^s(t,\mathbf{x})  +  b_{\mathbf{k}}^{s^\dagger} \bar{u}_{\mathbf{k}}^s(t,\mathbf{x}) \big].
	\end{align}

\subsection{Scalar modes}
For the scalar field modes, one needs to solve the curved space Klein-Gordon equation \eqref{eq:KGCurved} and identify the positive modes.
The scalar field $\phi$ can be expanded in the usual way as
	\begin{align}
	\phi&=\int d^3\mathbf{p}\big[ a_{\mathbf{p}}w_{\mathbf{p}}(t,\mathbf{x})  +  d_{\mathbf{p}}^\dagger w_{\mathbf{p}}^*(t,\mathbf{x}) \big]
	\end{align}		
and because of the homogeneity of the spatial sections, the positive mode solutions are separable into time and space parts	
\begin{align}
	w_\mb p(t,\mb x)=\frac{e^{i\mb p\cdot \mb x}}{(2\pi)^{3/2}a(t)}\chi_p(t),
\end{align}
where $p:=|\mb p|$. The positive mode is found by solving the ensuing differential equation for $\chi_p(t)$ and recognizing the positive frequency modes in the usual way \citep{Birrell_Davies}.

The normalized solutions $\chi_p(t)$ were obtained in \citep{Lankinen_Vilja:2018b} for a massive scalar field mode with an arbitrary coupling to gravity in the rest frame of the particle $\mb p=\mb 0$. The rest frame was used in order to get an exact solution for the decay rate in \citep{Lankinen_Vilja:2018b} as well as in previous studies \citep{Lankinen_Vilja:2018a,Lankinen_Vilja:2017b}. In what follows, it turns out that in order to find an exact solution for the transition amplitude, we will also in this paper need to restrict to the rest frame of the decaying particle. Hence the solution obtained in the rest frame is sufficient for our purposes. The solution for the massive scalar mode from \citep{Lankinen_Vilja:2018b} reads as
	\begin{align}
	\chi_{p=0}(\eta)=\sqrt{\frac{\pi\eta}{2(2+n')}}e^{-\frac{i\pi}{4}(1-2\alpha)}\HankelH_\alpha 	 \Big(\frac{2b'm\eta^{(2+n')}/2}{2+n'}\Big),
	\end{align} 
where $\eta$ denotes the conformal time and index $\alpha$ is defined as
	\begin{align}
	\alpha:=\frac{\sqrt{1-n'(n'-2)(6\xi-1)}}{2+n'}.
	\end{align}	
	
At this point we have to reconcile the issue that different scale factors with different time variables were used for solving the Dirac equation above and what was obtained for the solution of the scalar particle. Since we are using a scale factor $a(t)=bt^n$ and in \citep{Lankinen_Vilja:2018b} the scale factor was given in conformal time as $a(\eta)=b'\eta^{n'/2}$, we must find the relation between the primed and unprimed variables. The coordinate time can be transformed into coordinate time using the usual relation $dt=a(\eta)d\eta$. Comparing the scale factors in coordinate time gives the relations between the primed and unprimed variables,
	\begin{align}
	n=\frac{n'}{2+n'},\quad b=b'\Big(\frac{2+n'}{2b'}\Big)^{\frac{n'}{2+n'}}.
	\end{align}
	
In terms of the standard coordinate time $t$ and the unprimed variables, the scalar field mode reads as
	\begin{align}\label{eq:ScalarMode}
	\chi_{p=0}(t)=\sqrt{\frac{\pi t^{1-n}}{4b}}e^{-\frac{i\pi}{4}(1-2\alpha)}\HankelH_\alpha (mt),
	\end{align}  
and the index $\alpha$ as 	
	\begin{align}
	\alpha=\frac{\sqrt{(1-n)^2-4n(2n-1)(6\xi-1)}}{2}.
	\end{align}	
This completes the necessary background formalism.

\section{Transition amplitude in curved space}\label{Sec:4}
Having established the proper formalism, we turn to the calculation of the total transition amplitude and the decay rate. In this section we will explicitly calculate the transition amplitudes corresponding to the two Feynman diagrams of Fig. \ref{fig:1}, verify the transition amplitude has a correct form in the Minkowskian limit and discuss about convergence issues near the spacetime singularity. We will end this section with a calculation of the exact decay rate for conformally coupled massive scalars.

\subsection{Total transition amplitude}
The matrix element corresponding to the diagram (a) can be written with the positive field modes as
	\begin{align}\nonumber
&\braket{\mathrm{out}, 1^\psi_{\mathbf{k_1}} 1^{\psibar}_{\mathbf{k_2}} |A| 1^\phi_{\mathbf{p}}, \mathrm{out}}\\
&=\lim_{\beta\to 0^+}\int d^4x  e^{-\beta t} \sqrt{-g}\, \ubar^s_{\mathbf{k_2}}(t,\mathbf{x}) v^{s'}_{\mathbf{k_1}}(t,\mathbf{x}) w_{\mathbf{p}}(t,\mathbf{x}).		
	\end{align}
For diagram (b), the matrix element is obtained by substituting the scalar mode with its complex conjugate $w_{\mb p}(t,\mb x)\rightarrow w_{\mb p}(t,\mb x)^*$. Upon inserting the positive field modes given in Sec. \ref{Sec:2}, this integral can further be simplified by taking the limit inside the integral by introducing a time cutoff at $t=T$. Moreover, the spatial integration can readily be done resulting in taking out the delta function $\delta(\mathbf{p-k_1-k_2})$ for diagram (a) or $\delta(\mathbf{p+k_1+k_2})$ in case of diagram (b). These correspond to the usual conservation of three-momentum.
Hence, in the end, after also performing the $\mb k_2$ integration, one is left with the amplitude squared corresponding to diagram (a) as
	\begin{align}\nonumber
	&|\braket{\mathrm{out}, 1^\psi_{\mathbf{k}} 1^{\psibar}_{\mathbf{p-k}} |A| 1^\phi_{\mathbf{p}}, \mathrm{out}}|^2=\frac{1}{8(2\pi)^3 |\mb k| |\mb{p-k}|}\\
	&\times \Bigg|\ubar(\mathbf{p}-\mathbf{k},s) v(\mathbf{k},s') \int_{T_0}^T \frac{\chi_p(t)}{a(t)}e^{\frac{i(k+|\mb{p-k})}{b(1-n)}t^{1-n}} dt\Bigg|^2
	\end{align}
and similarly for diagram (b) with the substitution of $\chi_p(t)\rightarrow \chi_p(t)^*$ and $\mb p -\mb k \rightarrow -\mb p-\mb k$. 

To proceed ,we switch to the rest frame of the decaying particle where, $\mathbf{p}=\mathbf{0}$. The reasons for this are twofold. First, we only know the massive scalar field mode for a general power-law expansion in its rest frame. Second, by taking $\mathbf{p}=\mathbf{0}$ the integrand simplifies to a form where we can use the properties of distributions to obtain an exact result.  We begin by calculating the spin sum in the rest frame giving
	\begin{align}
	\sum_{s,s'}|\ubar(-\mathbf{k},s) v(\mathbf{k},s')|^2=8k^2,
	\end{align}
where $k=|\mathbf{k}|$. Because the spin sum is proportional to the square of the momentum vector length, we can combine the $k$-integration of both diagrams into
	\begin{align}
	w^{\mathrm{tot}}=\frac{h^2}{2\pi^2}\int_{-\infty}^\infty k^2dk \Bigg|\int_{T_0}^T \frac{\chi_{p=0}(t)}{a(t)}e^{\frac{2ik}{b(1-n)}t^{1-n}} dt \Bigg|^2
	\end{align}
by a change of variables $k\rightarrow -k$ in amplitude squared of diagram (b).	
Upon inserting the scalar field mode \eqref{eq:ScalarMode} we obtain the following integral
	\begin{align}\label{eq:wtot2}
	w^{\mathrm{tot}}\!=\frac{h^2}{16\pi b^3}\int_{-\infty}^\infty k^2dk \Bigg|\int_{T_0}^T t^{\frac{1-3n}{2}} H^{(2)}_\alpha (mt) e^{\frac{2ik}{b(1-n)}t^{1-n}} dt \Bigg|^2
	\end{align}	
The above integral can be written as a three dimensional integral which allows the use of the distribution identity
	\begin{align}\label{Id:Deltaprime}
	\int_{-\infty}^\infty k^2 e^{ik(y-x)}dk=-2\pi\delta''(y-x)
	\end{align}
on the $k$-integration, which reduces Eq. \eqref{eq:wtot2} into a two-dimensional integral. The remaining integrations can be performed in a symmetric fashion using derivative identities of distributions\footnote{For any distribution $f$ and test function $\varphi$ with a compact support it holds that $\int f'\varphi dx=-\int \varphi'f dx$.}. Using also the fact that  $\delta'(y-x)=-\delta'(x-y)$ we can switch the order of integration to obtain an integral which has two derivative terms. In the end, by performing the remaining delta integration, we arrive at the exact form for the total transition probability as
	\begin{align}\label{eq:TransitionProb}
	w^{\mathrm{tot}}=\frac{h^2}{32}\int_{mt_0}^{mt} \bigg|\frac{d}{du}\Big(u^{\frac{1-n}{2}}H^{(2)}_\alpha(u) \Big)  \bigg|^2 u^n du,
	\end{align}
where we have changed to a dimensionless variable $u=m(\frac{b(1-n)s}{2})^{1/(1-n)}$. Because of the absolute value squared, the realness of Eq. \eqref{eq:TransitionProb} is manifest. Although derived using the real scalar field modes, the total transition probability using the complex field  is just Eq. \eqref{eq:TransitionProb} multiplied by a factor of two.

The equation \eqref{eq:TransitionProb} is our main result of this paper. It is an exact formula for the transition probability from which the decay rate may be obtained. The integration can be performed exactly resulting in an extremely long and complicated expression involving hypergeometric and Bessel functions. This expression is of no illuminating value and we do not present it here. 

In its current form, Eq. \eqref{eq:TransitionProb} gives the mean decay rate when divided by the time $t$. Of interest besides this is also on the form of a differential decay rate which can readily be obtained as the integrand of the transition probability. If we make this interpretation the differential decay rate is given as
	\begin{align}\label{eq:GammaDiff}
	\Gamma_{\psi}^{\mathrm{diff}}=\frac{h^2 t^n}{32}\bigg|\frac{d}{dt}\Big(t^{\frac{1-n}{2}}H^{(2)}_\alpha(mt) \Big)  \bigg|^2.
	\end{align}

\subsection{Minkowskian limit}

As with any theory in physics, a generalized theory should reproduce the known results in the more specialized situation. Therefore a decay rate in an expanding spacetime should yield the flat spacetime results when the appropriate limit is taken. This limit is given when the spacetime is static, i.e., the scale factor $a(t)=1$. This is achieved for the value $n=0$ and, although not needed in the final equation \eqref{eq:TransitionProb}, one may take $b=1$ since it is only a scaling factor. With the value $n=0$, the parameter $\alpha=1/2$ and the transition probability reads as
	\begin{align}\label{eq:MinkowskianTransProb}
	w^{\mathrm{tot}}_{\mathrm{Mink}}=\frac{h^2}{32}\int_{mt_0}^{mt} \bigg|\frac{d}{du}\Big(u^{\frac{1}{2}}H^{(2)}_{1/2}(u) \Big)  \bigg|^2 du.
	\end{align}
Because $H^{(2)}_{1/2}(u)=i\sqrt{2/ (\pi u)}e^{-iu}$, we see immediately that the integrand in \eqref{eq:MinkowskianTransProb} is equal to $2/\pi$ and the transition amplitude reduces to
	\begin{align}\label{eq:MinkProb2}
	w^{\mathrm{tot}}_{\mathrm{Mink}}=\frac{h^2m}{16\pi}(t-t_0).
	\end{align}
From this, the decay rate is obtained in the usual way by dividing the transiton probability by the infinite time $t$. Thereby dividing equation \eqref{eq:MinkProb2} by $\Delta t:=t-t_0$, we obtain the Minkowskian decay rate
	\begin{align}
	\Gamma_{\mathrm{Mink}}=\frac{h^2m}{16\pi}
	\end{align}
which corresponds to the decay rate calculated in the added-up method using only Minkowskian plane waves. This consistency serves as a validation of our procedure.

\subsection{Convergence near spacetime singularity}
Equation \eqref{eq:TransitionProb} gives the transition probability beginning from some time $t_0$ up to time $t$, assuming that the total transition probability $w^{\rm{tot}}\ll 1$. Although e.g., in post-inflationary scenarios the time $t_0$ truly differs from zero, in some instances it may be necessary to take the limit $t_0\to 0$. Whether for mathematical reasons to simplify the integration or to investigate the decay process all the way from singularity. It is therefore illuminative to study also the zero time limit of the transition probability.

The convergence of the integral of Eq. \eqref{eq:TransitionProb} in the limit where $t_0\rightarrow 0$ can be investigated in a simple fashion. First, the Hankel function of the second kind has the following limiting form when the argument approaches zero: $H^{(2)}_\alpha(u)\sim i\Gamma(\alpha) (u/2)^{-\alpha}/\pi$. Second, requiring that the behavior of the power of the variable $u$ is greater than $-1$, to avoid  singular behaviour, we can restrict the values of $\alpha$ for which we obtain a finite result. The condition for which the transition probability is finite is given by the requirement $\alpha<0$. This implies that the integral diverges for all values of $n$ and $\xi$ when $t_0\rightarrow 0$ is taken.

There is one value, namely $\alpha=(1-n)/2$, however for which the limit is finite. The reason for this is that for this value, the power of the limiting form of the Hankel function exactly cancels out the power of the prefactor leaving only $u^n$ to be integrated. Hence, the transition probability Eq. \eqref{eq:TransitionProb} converges near the spacetime singularity when the following condition holds,
	\begin{align}
	\frac{\sqrt{(1-n)^2-4n(2n-1)(6\xi-1)}}{2}=\frac{1-n}{2}.
	\end{align}

This condition is valid for Minkowski limit $n=0$ as well as for radiation dominated universe $n=1/2$ for any value of the coupling. But the most peculiar feature of it is that the equality holds for conformal and only conformal coupling $\xi=1/6$ and for any $n\in (0,1)$. Now that we know the integral of Eq. \eqref{eq:TransitionProb} is convergent for conformal coupling at the spacetime singularity, we can calculate it explicitly.

\subsection{Conformally coupled massive particles}
As noted, the transition probability \eqref{eq:TransitionProb} is finite for all $n\in [0,1)$ in the limit $t_0\rightarrow 0$ only for conformally coupled massive particles. For a conformally coupled massive scalar particle, the decay probability reads as
	\begin{align}\nonumber
	w^{\mathrm{tot}}&=\frac{h^2}{32}\int_{0}^{mt} \bigg|\frac{d}{du}\Big(u^{\frac{1-n}{2}}H^{(2)}_{\frac{1-n}{2}}(u) \Big)  \bigg|^2 u^n du\\
	&=\frac{h^2}{32}\int_0^{mt} u H^{(2)}_{-\frac{1+n}{2}} H^{(1)}_{-\frac{1+n}{2}}du,
	\end{align}
where, to get to the second line, we have used the derivative identity $[u^\alpha H^{(2)}_\alpha]'=u^\alpha H^{(2)}_{\alpha-1}$. This integral can be integrated exactly to yield
	\begin{align}\nonumber
	w^{\mathrm{tot}}&=\frac{h^2}{64}\{(mt)^2[J_{-\frac{1+n}{2}}(mt)^2-J_{-\frac{3+n}{2}}(mt)J_{\frac{1-n}{2}}(mt)\\\nonumber
	&+Y_{-\frac{1+n}{2}}(mt)^2-Y_{-\frac{3+n}{2}}(mt)Y_{\frac{1-n}{2}}(mt)]\\
	&+\frac{2(1+n)\tan(n\pi/2)}{\pi} \},
	\end{align}
where the last constant term arises from the lower limit of the integration. For further insight, we take a look at the asymptotic behaviour of this transition amplitude when the time is large. Given by the leading terms in the asymptotic expansion, the transition probability is given by
	\begin{align}
	w^{\mathrm{tot}}&\sim\frac{h^2m}{16\pi}\Big( t +\frac{(1+n)}{2m}\tan(\frac{n\pi}{2})\Big).
	\end{align}
	
We see that it is linear in time $t$, which would prompt us to divide the transition probability by the time $t$ to obtain the decay rate in the familiar fashion of Minkowski space field theory. A complication however arises from the constant term reflecting the fact that in curved space one needs to be more careful concerning usual Minkowski space procedures. To deal with the constant term we adopt the same procedure as was introduced in \citep{Audretsch_Spangehl:1985} where this constant term is divided by a time $t_{grav}$, known as time of gravitational influence. This can be defined as $t_{grav}:=t_f-t_i$, where $t_i$ indicates the time when the gravitational field begins its influence and $t_f$ its end. With this definition the asymptotic decay rate is given by
	\begin{align}\label{eq:GammaPsi}
	\Gamma_{\psi}&\sim\frac{h^2m}{16\pi}\Big(1 +\frac{(1+n)\tan(n\pi/2)}{2mt_{grav}}\Big).
	\end{align}
	
It is now seen that asymptotically the decay rate obtains a gravitational correction term to the Minkowskian decay rate. This term is positive implying, in the case of conformal coupled particles, that the effect of the gravitational field is to enhance the decay rate. With Eq. \eqref{eq:GammaPsi}  we are now in a position to compare this to the process where a massive scalar decays into two massless scalars.

\section{Comparison with scalar to scalar decay}\label{Sec:5}

In this section we will mainly be concerned in comparing the results of this paper into studies where scalar to scalar decay was concerned using also the added-up method \citep{Lankinen_Vilja:2018b,Lankinen_Vilja:2018a,Lankinen_Vilja:2017b}. This allows a direct comparison due to the same methods of calculations being used.

The asymptotic decay rate for a massive scalar $\phi$ decaying into two conformally coupled massless scalars $\chi$ was given in \citep{Lankinen_Vilja:2018b} as
	\begin{align}\label{eq:GammaPhi}
	\Gamma_{\phi}\sim \frac{\lambda^2}{16m\pi}\Big( 1-\frac{(1-n)\cot[(1-n)\pi/2]}{4mt_{grav}}\Big),
	\end{align}
where $\lambda$ is the coupling constant. This expression was also derived in the limit of $t_0\to 0$ making the comparison valid. Comparing the expressions \eqref{eq:GammaPsi} and \eqref{eq:GammaPhi} a major difference is immediately seen corresponding to the sign of the relative correction term. For a decay into scalars, the sign is negative, implying that the effect of a gravitational field is to diminish the rate of decay. On the contrary, in fermionic decay the sign is positive and the effect of the gravitational field is to enhance the decay. Both relative correction terms are increasing functions of the parameter $n$ on the interval $n\in [0,1)$, so the gravitational correction term increases in magnitude as $n$ increases. This means that the decay into fermions increases and decay into scalars diminishes as the parameter $n$ increases. 

\begin{figure}[H]
	\centering
	\includegraphics[width=1.0\columnwidth]{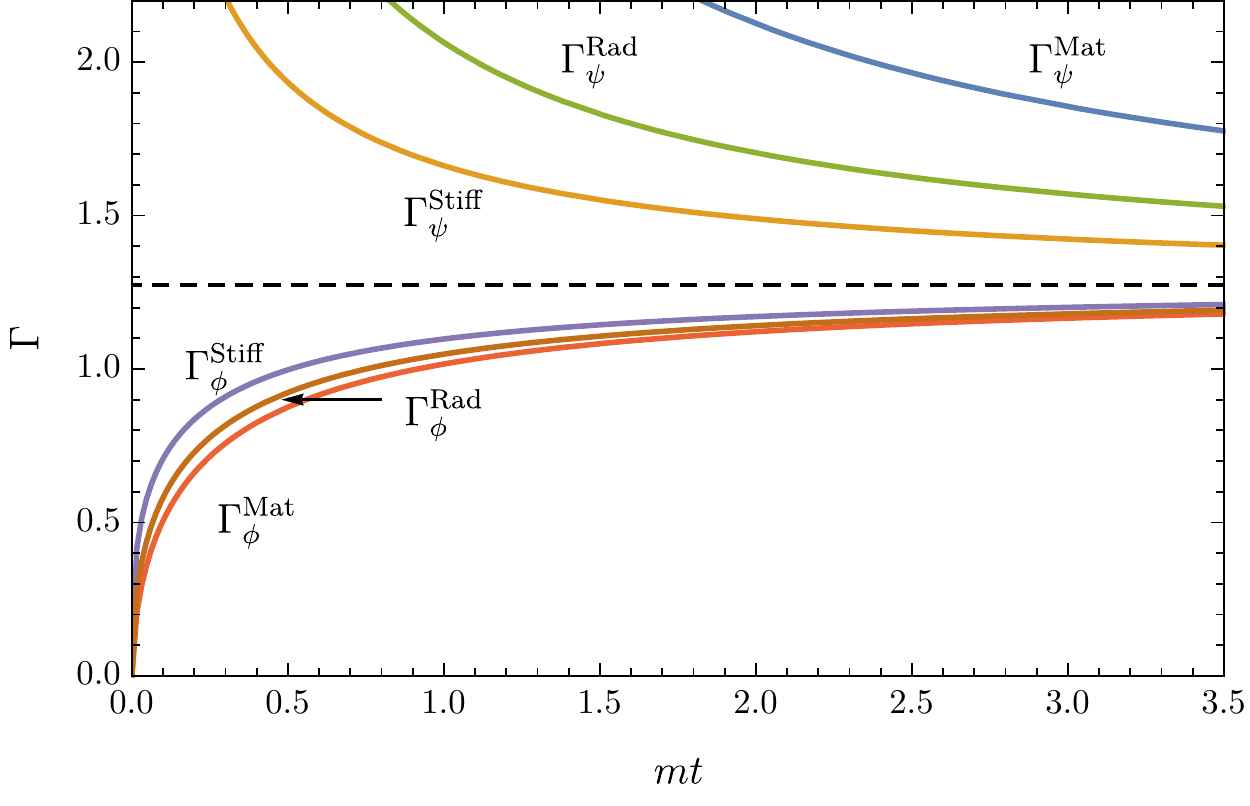}
	\caption{Exact mean decay rate $\Gamma_\phi$ for the scalar channel in units of $h^2m/64$ and $\Gamma_\psi$ for the fermion channel in units of $\lambda^2/64m$ as a function of $mt$ for radiation, matter and stiff matter dominated universes. The dashed line corresponds to asymptote at $4/\pi$  for all decay rates.}\label{fig:2} 
	\end{figure}

Although the ordering of these decay rates is most prominently seen in the asymptotic decay rates, it cannot be inferred from these that the decay rates are so ordered for all time. To investigate the small time behaviour, one needs to take a look at the exact decay rates. This is done in Fig. \ref{fig:2}  for three cosmologically interesting situations, namely universes dominated by stiff matter, radiation and ordinary matter. From Fig. \ref{fig:2} we see that this ordering of the decay rates is the same for these universes for all time and not just asymptotically.
Finally, it should be noted that in practical situations $t_{grav}$ is usually much longer than inverse of mass, when these corrections are in practice small. These relative correction terms might not, however, be neglegted altogether, because for $t\sim m$ the exact equation \eqref{eq:MinkowskianTransProb} divided by $t$ should be used instead.

As was noted, the integral in Eq. \eqref{eq:TransitionProb} diverges when $t_0$ approaches the spacetime singularity. When looking at the behavior of the differential decay rate \eqref{eq:GammaDiff}, interpreted as the integrand of \eqref{eq:MinkowskianTransProb}, we notice that it behaves like $t^{-2\alpha-1}$ in the vicinity of the spacetime singularity. On the other hand, the scalar to scalar differential rate obtained in \citep{Lankinen_Vilja:2018b},
	\begin{align}
	\Gamma^{\textrm{diff}}_{\psi}=\frac{\lambda^2t}{32} \big| H^{(2)}_\alpha(mt)\big|^2
	\end{align}
 behaves like $t^{-2\alpha+1}$ near the singularity. These observations together imply that near the singularity, as long as $2\alpha+1>2\alpha-1$, the decay channel into fermions is dominant over the decay channel into scalars. However, this inequality is always true implying that in the very early Universe, the decay channel into fermions is always the dominant one to the scalar channel no matter what is the scalar coupling or universe matter content. It must be stressed, however, that this holds only near the singularity and the exact time when it is true is not evident from this analysis because the decay rates are differently proportional to the couplings and masses in the prefactors which affect the decay rate.

\section{Discussion}\label{Sec:6} 
Our study of scalar particle decay in the early Universe has presented us with novel features concerning the effect of gravitation on the decay process. Not only does the curved background modify the Minkowskian results, but the effect is in some cases opposite for fermionic and scalar channels.

We wish to emphasize three key elements of this paper:
\begin{enumerate}[label=(\alph*)]
\item We have derived a curved space generalization for a massive scalar to decay into two massless  fermions.
\item For conformally coupled scalars, the decay into fermionic channel is enhanced and decay into scalar channel diminished by the curved background.
\item It is found that in the very early Universe, the fermionic decay channel is dominant to the scalar channel.
\end{enumerate}

These two last points bear some further elaboration.
For conformally coupled scalars, the integration from the singularity introduces an additive constant to the decay rate. In the paper \citep{Lankinen_Vilja:2018a} a proposition was put forward in the case of a scalar to scalar decay that the faster the Universe is expanding, the smaller is the decay rate into scalars. We may elaborate this point by speculating within the same framework, i.e., by taking a look at the Hubble parameter $H=\dot{a}/a=n/t$. The Hubble parameter shows us that as $n$ increases, the Universe expands relatively faster. In the case of a fermionic decay channel, the faster the Universe is expanding the larger is the decay rate. We may speculate the reason behind this from a statistical point of view. As the Universe expands faster, more and more states are becoming available for the fermions to occupy. For bosons, on the other hand, this same expansion reduces the Bose enhancement thereby diminishing the decay. This would statistically explain the observed phenomena, but further investigations into this feature are surely needed.

As was noted, independent of the matter content of the Universe or the coupling of the massive scalar, the fermionic decay channel is the dominant one in the very early Universe. How close this time should be to the singularity for the fermionic channel to be the dominant one cannot be inferred from the analysis made in Sec. \ref{Sec:4}. For times farther away from the singularity, it cannot be said that for any fixed $n$ the decay into fermions is more probable because the prefactors on the decay rates differ; the scalar to scalar is proportional to $m^{-1}$ while scalar to fermion is proportional to the mass $m$ with different couplings. Furthermore our study has been restricted to massless fermions in the fermionic decay channel. The main reason for this has been the fact that in order to apply the added-up method and to obtain a meaningful decay rate interpretation in curved space this needs to be done. For fermions this restriction is, however, less severe when considering times before electroweak phase transition because then fermions are usually massless as in the Standard Model. If the time when the fermion channel is dominant over the scalar channel is before this phase transition, then it could be inferred that the decay into Standard Model fermions is more preferred than massless scalars. Which brings up another point. 

The analysis for scalar to scalar decay in \citep{Lankinen_Vilja:2018a} was done using massless scalars as decay products. The inference that the fermion decay channel is dominant to the scalar channel only applies when the product scalar particles are massless, although one could argue that for very light scalars this is still valid. The case where the decay products are also massive cannot be straightforwardly calculated using the added-up method and it cannot be said if the fermionic channel is indeed the dominant one in all cases. The decay into massive scalars has been considered in \citep{Herring:2018} using the methods introduced in \citep{Boyanovsky_Holman:2011,Lello_Boyanovsky_Holman:2013} for post inflationary cosmology where the authors obtained similar type of results indicating slower decay for scalars. It would be a worthwhile study to see whether similar types of results are obtained for a fermionic channel as well.

\section{Conclusions}\label{Sec:7} 

Particle decay rates in the early Universe play an ever increasing role in modern cosmology. In this paper we have used quantum field theory in curved spacetime and the added-up probability method to calculate the decay rate of a $\phi\to\psi\psibar$ process in curved spacetime. We have found that the Minkowskian decay rates are modified in the early Universe by the presence of a gravitational field and that the fermionic decay channel of a scalar is dominant to a scalar decay channel in the very early Universe.

This modification of decay rates should be taken into account when considering  with more precision early universe processes involving scalar decay scenarios. For fermionic decay of a scalar, possible scenarios involve e.g., fermionic preheating scenarios. Also, as cosmological data and measurements become increasingly more accurate, it may be necessary in the future to include the effects of curved space also in particle decay rates.

\begin{acknowledgments}
J.L. would like to acknowledge the financial support from the University of Turku Graduate School (UTUGS).
\end{acknowledgments}

\end{document}